\documentclass[prb,twocolumn]{revtex4}

\usepackage{psfrag}
\usepackage{times}
\usepackage{amsmath}
\usepackage{graphicx}

\newcommand{\dd}{\mbox{d}}

\begin{document}

\title{Percolation, Bose-Einstein Condensation, and String
Proliferation}
\author{Adriaan M. J. Schakel} 
\affiliation{Low Temperature Laboratory, Helsinki University of Technology,
P.O. Box 2200, FIN-02015 HUT, Finland \\ and \\ Institut f\"ur
Theoretische Physik, Freie Universit\"at Berlin, Arnimallee 14, 14195
Berlin, Germany }

\date{\today}

\begin{abstract}
The close analogy between cluster percolation and string proliferation
in the context of critical phenomena is studied.  Like clusters in
percolation theory, closed strings, which can be either
finite-temperature worldlines or topological line defects, are described
by a distribution parametrized by only two exponents.  On approaching
the critical point, the string tension vanishes, and the loops
proliferate thereby signalling the onset of Bose-Einstein condensation
(in case of worldlines) or the disordering of the ordered state (in case
of vortices).  The ideal Bose gas with modified energy spectrum is used
as a stepping stone to derive general expressions for the critical
exponents in terms of the two exponents parameterizing the loop
distribution near criticality.
\end{abstract}
\maketitle

\section{Introduction}
The quest for a geometrical description of phase transitions has a long
history going back to ideas first put forward by Onsager \cite{Onsager}
in the context of the $\lambda$-transition in liquid $^4$He.  The
relevant geometrical objects in this transition are topological line
defects.  The description envisaged by Onsager is one entirely in terms
of these 1-dimensional objects, with their geometrical properties such
as fractal dimension and configurational entropy.  The phase transition
is characterized in this picture by a fundamental change in the typical
string size.  Whereas in the superfluid phase only finite strings are
present, at the critical point infinite strings appear---similar to the
sudden appearance of a percolating cluster in percolation phenomena at
criticality.

The analogy between percolation of clusters and proliferation of strings
has been noted by various authors
\cite{Vachaspati,Bradleyetal,HindmarshStrobl,Akao}.  As we will see, it
derives from a similarity in the cluster and string distribution.  Both
have the same form containing two factors, one related to the entropy of
a given cluster or string configuration, and the other related to the
Boltzmann weight assigned to the configuration.  Close to criticality,
each of these factors is parametrized by a single exponent.  One
specifies the algebraic behavior of the distribution at criticality,
while the other describes how the Boltzmann factor tends to unity upon
approaching the critical point.  Physically, the unity of the Boltzmann
factor implies that clusters or strings can grow and, thus, gain
configurational entropy without energy cost.  When the Boltzmann factor
is not unity, the clusters and strings are exponentially suppressed.  In
the context of strings, the transition between a phase consisting of
finite strings only and one having infinite strings, is called a
Hagedorn transition.  We will refer to the appearance of infinite
strings as proliferation.

A central position in our arguments is taken by an ideal Bose gas with
modified energy spectrum.  The reason is that although noninteracting,
the model has nontrivial critical exponents, which are known exactly,
while at the same time it can also be mapped onto a loop gas of
worldlines in an exact way.  The worldlines form closed loops because in
the absence of external fields and sources they cannot terminate in the
system.  The map onto the loop gas allows us to connect the critical
exponents of the phase transition to the two exponents parameterizing
the worldline loop distribution near criticality.  Using general scaling
relations, these results can then be generalized to interacting loop
gases representing statistical models.  Each universality class is
defined by a loop distribution with specific values for the two
exponents from which all the critical exponents follow.

To close the circle, the resulting loop gas description of critical
phenomena can also be applied to phase transitions involving
(interacting) vortex loops---like the $\lambda$-transition in liquid
$^4$He.  As for worldlines, vortices cannot terminate in the system and,
therefore, form closed loops too.  When the word string is used in the
following, the two different 1-dimensional objects worldlines and
vortices should be kept in mind.

As an aside, we believe that the theory of vertex loop gases discussed
here has also a bearing on homogeneous superfluid turbulence.

The paper is organized as follows.  In the next section those essentials
of percolation theory are recalled which later on in the paper become
important to establish the connection with the proliferation of strings.
In Sec.\ \ref{sec:BEC}, Bose-Einstein condensation in an ideal Bose gas
with a modified energy spectrum is studied from the perspective of
worldlines.  In Sec.\ \ref{sec:String}, numerical work on random string
networks is discussed in relation to uncorrelated percolation, followed
by a discussion of correlated percolation in Sec.\ \ref{corper}.  In
Sec.\ \ref{sec:LG}, thermal phase transitions are formulated in terms of
proliferating worldlines.  In Sec.\ \ref{sec:VL}, a similar formulation
is discussed for phase transitions involving proliferating vortices.
The paper ends with conclusions in Sec.\ \ref{sec:concl}.

\section{Review of Site Percolation}
In this section, we briefly recall some basic aspects of percolation
theory \cite{StauferAharony} which are important for our purposes.
Consider a lattice, with each lattice site being randomly occupied with
a probability $p$, say, independent of its neighbors.  Island of
next-neighboring occupied sites form clusters.  The properties of these
clusters as a function of $p$ form the subject of percolation
theory---or more precisely, of {\it uncorrelated} percolation theory
because the occupations of different lattice sites are uncorrelated.  As
$p$ increases, clusters become bigger, and at some critical value
$p_{\rm c}$, a cluster spanning the entire lattice first appears.  Near
this percolation threshold, various quantities show power-law behavior.
As for thermal critical phenomena, this behavior can be characterized by
a set of critical exponents of which two are independent.

Of interest to us is the so-called percolation strength $P(p) \sim (p-
p_{\rm c})^{\beta_{\rm per}}$ defined for $p > p_{\rm c}$ as the
probability that a randomly chosen site belongs to the percolating
cluster, the average cluster size $S(p) \sim |p_{\rm c} - p|^{-
\gamma_{\rm per}}$, and the correlation length $\xi(p) \sim |p_{\rm c} -
p|^{- \nu}$.  We give some quantities a subscript ``per'' to avoid
confusion later on, where similar, but different quantities appear.  The
percolation strength is analogous to the magnetization in spin models,
while the average cluster size corresponds to the magnetic
susceptibility.

The number density of clusters having size $s$, i.e., with $s$ occupied
sites, is assumed to be distributed according to
\begin{equation} 
\label{perdis}
l_s (p) \propto s^{- \tau} \exp(- c s),
\end{equation} 
where the coefficient $c$ vanishes with an exponent $1/\sigma$ when the
percolation threshold $p_{\rm c}$ is approached from below $c \propto
(p_{\rm c} - p)^{1/\sigma}$.  At criticality, the cluster distribution
becomes $l_s (p_{\rm c}) \propto s^{- \tau}$ and falls off
algebraically.  This factor measures the configurational entropy of
clusters, while the exponential is a Boltzmann factor.  Together with
the so-called Fisher exponent $\tau$, the exponent $\sigma$ determines
the critical exponents, such as $\beta_{\rm per}, \gamma_{\rm per}$ and
$\nu$ through scaling relations.

In terms of the cluster distribution $l_s$, the average cluster size is
given by
\begin{equation}
\label{S} 
S(p) = \frac{\sum_s s^2 l_s(p)}{\sum_s s l_s(p)},
\end{equation} 
where only finite clusters are included in the sum.  To understand that
$S$ thus defined indeed is a measure for the average cluster size, note
that the combination $s l_s(p)$ in the denominator is the probability
that a randomly chosen site belongs to a cluster of size $s$.  When
summed over $s$, this gives the probability that the randomly chosen
site belongs to a cluster of arbitrary (finite) size.  The ratio $s
l_s(p)/\sum_s s l_s(p)$ is therefore the probability that the cluster to
which the randomly chosen site belongs is of size $s$.  When this ratio
is multiplied with the cluster size and summed over $s$, one obtains
indeed a measure of the average cluster size.  With the explicit
expression for $S$ one easily finds that
\begin{equation} 
\label{gammaper}
-\gamma_{\rm per}  = \frac{\tau-3}{\sigma}.
\end{equation} 
For later reference, recall that the susceptibility also equals the
integral over the correlation function $G_{\rm per}(x)$ of the system under
consideration.  Because of Eq.\ (\ref{S}) one finds for percolation
\begin{equation}
\label{sumGper} 
\sum_x \; G_{\rm per}(x) \propto \sum_s s^2 l_s(p),
\end{equation} 
where the sum $\sum_x$ is over all lattice sites.  Physically, $G_{\rm
per}(x)$ is the probability of finding at a distance $x$ from the origin
an occupied site belonging to the same cluster as the origin does.  At
criticality, it has in $d$ space dimensions the algebraic behavior
\begin{equation} 
\label{algebraic}
G_{\rm per}(x) \sim \frac{1}{x^{d-2 + \eta_{\rm per}}},
\end{equation} 
with $\eta_{\rm per}$ a critical exponent.  

To obtain a similar expression for the percolation strength as given in
Eq.\ (\ref{S}) for $S$, one uses the identity
\begin{equation} 
p = P(p) + \sum_s s l_s(p),
\end{equation} 
stating that an occupied site either belongs to the percolating cluster
or to a finite one.  Hence, for $p > p_{\rm c}$
\begin{equation}
\label{P} 
P(p) \sim  - \sum_s s l_s(p).
\end{equation} 
and
\begin{equation}
\label{betaper} 
\beta_{\rm per} = \frac{\tau-2}{\sigma}.
\end{equation} 
Below the percolation threshold, $P(p)=0$, so that the identity becomes
$p = \sum_s s l_s(p)$.

Given the analogy with spin systems, it should be possible to define a
partition function $Z$ from which the percolation strength
(magnetization) and the average cluster size (susceptibility) can be
obtained by differentiation.  Indeed, with the definition
\begin{equation} 
\label{perZ}
\ln(Z) \propto \sum_s l_s,
\end{equation} 
one finds
\begin{equation} 
\label{PS}
P = \frac{\partial \ln(Z)}{\partial c}, \quad S = \frac{\partial^2
\ln(Z)}{\partial c^2},
\end{equation} 
as required by the analogy.  Physically, $\ln(Z)$ denotes the total
number density of finite clusters.

The two critical exponents $\gamma_{\rm per}$ and $\beta_{\rm per}$ can
be related to the critical exponent $\alpha$ characterizing the
algebraic behavior of $\ln(Z) \sim |p_{\rm c} - p|^{2- \alpha}$ close to
the percolation threshold as follows
\begin{equation} 
\label{Rushbrooks}
2 - \alpha = \frac{\tau-1}{\sigma} = 2 \beta_{\rm per} + \gamma_{\rm per}.
\end{equation} 
On dimensional grounds, one expects that near the percolation threshold
$\ln (Z) \sim \xi^{-d} \sim |p - p_{\rm c}|^{d \nu}$, so that one
arrives at the hyperscaling relation
\begin{equation} 
\label{Josephson}
d \nu = 2 - \alpha,
\end{equation} 
involving the dimensionality $d$ of the lattice.  In the context of
thermal critical phenomena, relations (\ref{Rushbrooks}) and
(\ref{Josephson}) are known as Rushbrooks' and Josephson's scaling law,
respectively.  A hyperscaling relation in general breaks down at the
upper critical dimension.  Beyond the upper critical dimension $d_{\rm
u}$, the critical exponents are locked in their values at $d=d_{\rm u}$.
For percolation, as we will see below, the upper critical dimension is
$d_{\rm u}=6$.

Another scaling relation between the various critical exponents is
obtained from the expressions (\ref{sumGper}) and (\ref{algebraic})
involving the correlation function.  Using $\xi$ as lower cutoff in the
integral, one finds
\begin{equation}
\label{Fisher} 
\gamma_{\rm per} = \nu (2 - \eta_{\rm per}),
\end{equation} 
which in the context of thermal phase transitions is known Fisher's
scaling law.

\begin{figure}
\begin{center}
%\leavevmode
\psfrag{d}[t][t][1][0]{$d$}
\psfrag{t}[t][t][1][0]{$\tau$}
\psfrag{s}[t][t][1][0]{$\sigma$}
\includegraphics[width=6.0cm]{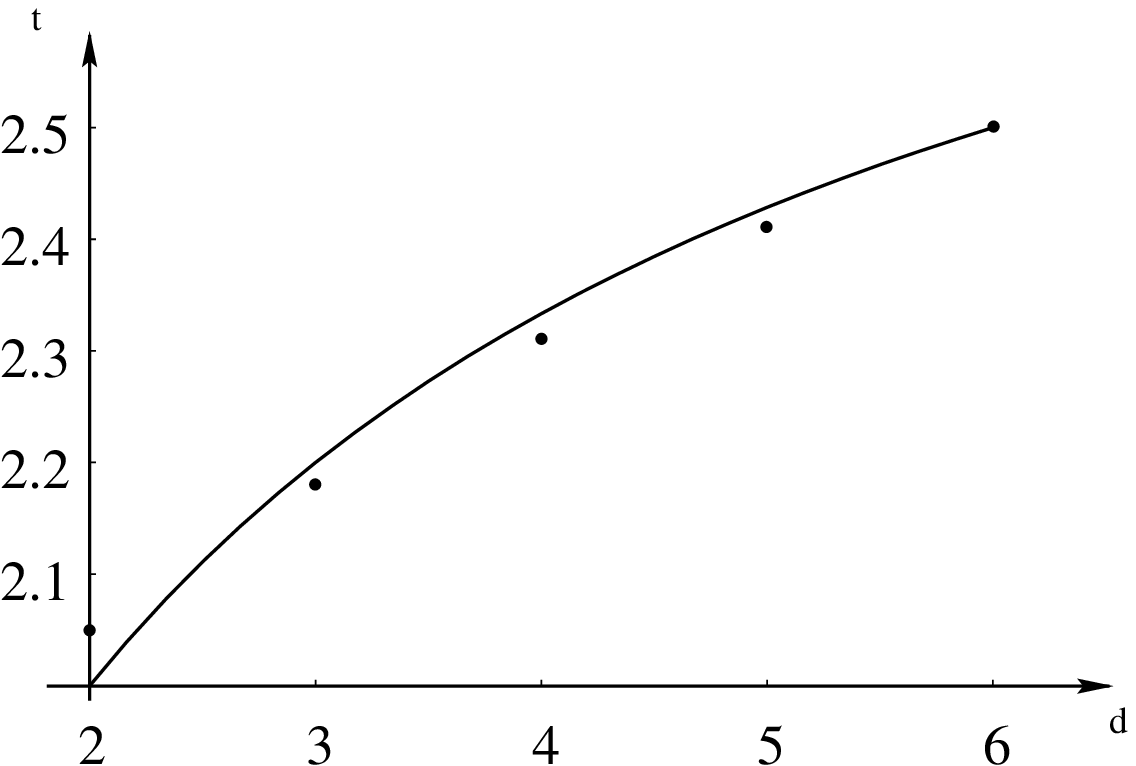}
\vspace{.7cm}

\hspace{-1.5em} \includegraphics[width=6.0cm]{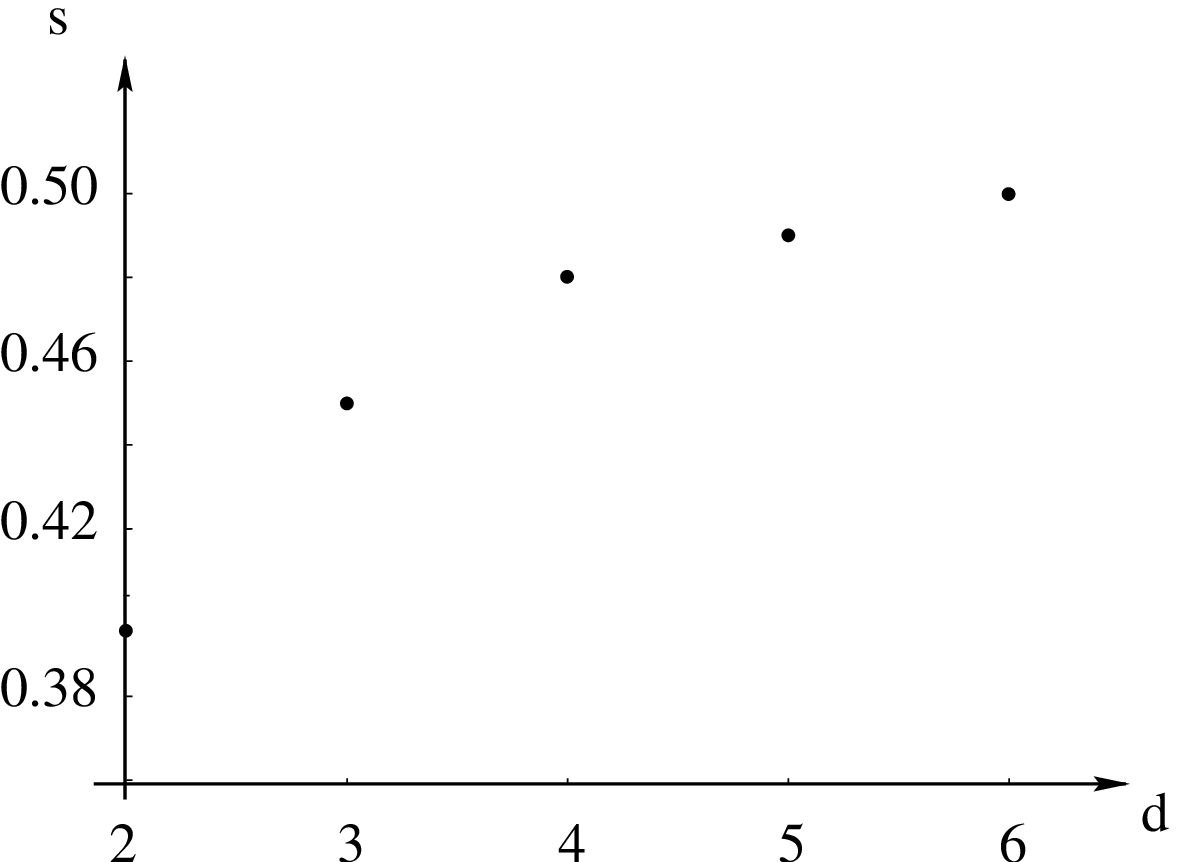}
\end{center}
\caption{The Fisher exponent $\tau$ and the exponent $\sigma$
parameterizing the cluster distribution of uncorrelated site percolation
as function of the dimensionality $d$ of the lattice.  The points
represent results \protect\cite{StauferAharony} obtained by lattice
simulations ($d=3,4,5$) and analytic methods ($d=2,6$), while the line
in the top panel is based on de Gennes' estimate (\protect\ref{deGennes})
which can be given for any dimension $2\leq d \leq 6$, not just integer
values.  \protect\label{fig:tau}}
\end{figure}

The correlation length exponent $\nu$ is known to have a geometric
meaning, being related to the Hausdorff, or fractal dimension $D$ at
criticality.  The latter can be defined through the average squared
distance between the sites of a cluster,
\begin{equation} 
R_s^2 = \frac{1}{s} \sum_{i=1}^s ({\bf x}_i - \bar{\bf x})^2 =
\frac{1}{2s^2} \sum_{i,j=1}^s ({\bf x}_i - {\bf x}_j)^2,
\end{equation} 
with $R_s$ the so-called radius of gyration, and $\bar{\bf x} = (1/s)
\sum_{i=1}^s {\bf x}_i$ the center of mass of the cluster with its sites
at ${\bf x}_i$, as
\begin{equation} 
\label{Hausdorff}
R_s \sim s^{1/D}
\end{equation} 
for large enough clusters.  A standard definition of the correlation
length in terms of the correlation function $G(x)$, with $x = |{\bf x}|$,
reads
\begin{equation}  
\label{def}
\xi^2 \propto \frac{\sum_x x^2 G(x)}{\sum_x  G(x)}.
\end{equation} 
It measures the average distance between two sites in the same cluster.
Because of the observation (\ref{sumGper}), this definition can be transcribed
in terms of the cluster distribution as
\begin{equation} 
\xi^2 \propto 2 \frac{\sum_s R_s^2 s^2 l_s(p)}{\sum_s s^2 l_s(p)}.
\end{equation} 
With Eq.\ (\ref{Hausdorff}), the relation between the correlation length
and the fractal dimension is then easily found to be
\begin{equation}
\label{nu} 
\frac{1}{D} = \nu \sigma.
\end{equation} 
Combined with this relation, the hyperscaling relation (\ref{Josephson})
leads to the well-known expression for the Fisher exponent $\tau$ in
terms of the fractal dimension:
\begin{equation}
\label{tau} 
\tau = \frac{d}{D} + 1.
\end{equation}  
Using a Flory-type \cite{KleinertPath} of argument known from polymer
physics and the theory of self-avoiding walks \cite{Flory}, de Gennes
\cite{deGennes} estimated the fractal dimension for uncorrelated site
percolation to be
\begin{equation} 
\label{deGennes}
D_{\rm dG} = (d+2)/2.
\end{equation}   
In Fig.\ \ref{fig:tau}, the Fisher exponent $\tau$ with this estimate is
compared with numerical ($d=3,4,5$) and analytic ($d=2,6$) results taken
from Ref.\ \cite{StauferAharony}.

Considering Eq.\ (\ref{sumGper}) at criticality with a long-distance
cutoff $x_{\rm max}= L$, corresponding to $s_{\rm max} = L^D$, we
obtain a relation between the fractal dimension and the critical
exponent $\eta_{\rm per}$ characterizing the algebraic behavior
(\ref{algebraic}) of the correlation function,
\begin{equation} 
\label{Deta}
D = \tfrac{1}{2} (d+2-\eta_{\rm per}).
\end{equation} 
The de Gennes' estimate (\ref{deGennes}) is equivalent to setting
$\eta_{\rm per}$ to zero here.

To determine the upper critical dimension it is prudent to consider
uncorrelated percolation on a Bethe lattice, which is exactly solvable.
Because of the absence of closed paths on such a lattice, it mimics an
ordinary lattice with high dimensionality, where one expects the same
critical behavior as in the upper critical dimension.  The cluster
distribution on a Bethe lattice is given by Eq.\ (\ref{perdis}) with
$\tau = 5/2$ and $\sigma = 1/2$, corresponding to the critical exponents
\begin{align} 
\alpha = -1,  \quad \beta_{\rm per} = 1, \quad \gamma_{\rm per} = 1, \quad
\nu = 1/2, \quad \eta_{\rm per}=0,
\end{align} 
and the fractal dimension $D=4$.  These values are consistent with the
scaling laws in $d=6$, which can thus be identified as the upper
critical dimension.

To summarize, percolation near criticality is specified by the two
exponents $\sigma$ and $\tau$ parameterizing the cluster distribution,
both of which are related to the fractal dimension $D$ via Eqs.\
(\ref{nu}) and (\ref{tau}).  Given the value of $\tau$ and $\sigma$, the
critical exponents can be obtained using scaling relations.
Specifically,
\begin{align} 
\label{perce}
\alpha & = 2 - \frac{\tau -1}{\sigma}, & \beta_{\rm per} &= \frac{\tau
-2}{\sigma}, & \gamma_{\rm per} & = \frac{3-\tau}{\sigma}, \nonumber \\
\eta_{\rm per} &= 2 + d \frac{\tau-3}{\tau-1}, & \nu & = \frac{\tau
-1}{d \sigma}, & D & = \frac{d}{\tau-1}.
\end{align}

\section{Ideal Bose Gases}
\label{sec:BEC}
In this section we wish to point out a close analogy between percolation
and Bose-Einstein condensation in an ideal Bose gas.  The partition
function $Z$ describing this system in $d$ space dimensions can be
written as an integral over momentum as \cite{Huang}:
\begin{equation} 
\label{kint}
\ln(Z) = - V \int \frac{\dd^d k}{(2 \pi)^d} \ln \left( 1 - {\rm
e}^{-E(k)/k_{\rm B} T} \right),
\end{equation} 
ignoring the $k=0$ contribution.  In Eq.\ (\ref{kint}), $V$ is the
volume of the system, while $E(k) = k^2/2m - \mu$ is the single-particle
spectrum, with $m$ the mass of the particles and $\mu$ the chemical
potential which accounts for a finite particle number density.  In
deriving this, the time coordinate was analytically continued to the
imaginary axis, where it becomes a cyclic variable taking values only in
the interval $[0, \hbar/k_{\rm B} T]$, with $k_{\rm B}$ Boltzmann's
constant.  After a Taylor expansion of the logarithm, the partition
function takes the familiar form of a fugacity [$=\exp(\mu/k_{\rm B}
T)$] series \cite{Huang}
\begin{equation} 
\label{explicit}
\ln(Z) =\frac{V}{\lambda^d} \sum_w l_w,
\end{equation} 
where
\begin{equation}
\label{loopdis} 
l_w(T)  = w^{- \tau} \exp(\mu w/k_{\rm B} T),
\end{equation} 
with $\tau = d/2+1$ and $\lambda = \hbar \sqrt{2 \pi/m k_{\rm B} T}$ the
de Broglie thermal wavelength.  

The correlation function $G(x)$ written as a sum over $w$ reads
\begin{equation}   \label{prop}
G(x) := \sum_w G_w(x) = \frac{1}{\lambda^d} \sum_w w \, l_w
\exp\left(-\frac{\pi x^2}{ \lambda^2 w}\right),
\end{equation} 
so that 
\begin{equation} 
\label{G0}
G(0) = \frac{1}{\lambda^d} \sum_w w \, l_w
\end{equation} 
and
\begin{equation} 
\label{sumG}
\int \dd^d x \, G(x) = \sum_w {\rm e}^{\mu w/k_{\rm B} T},
\end{equation} 
where the left side of the last equation denotes $\tilde{G}(0)$, with
$\tilde{G}(k)$ indicating the Fourier transform of $G(x)$.  Physically,
the correlation function $G_w(x)$ denotes the probability for a Brownian
random walk starting at the origin to end up at a distance $x$ from the
origin after $w$ steps, while $G_w(0)$ denotes the probability for the
random walk to return to its starting point after $w$ steps.  

As last quantity we need the correlation length $\xi$ which is given by
$\xi^2 = \lambda^2 k_{\rm B} T/(-4 \pi \mu)$, i.e., $\xi \propto
\mu^{-1/2}$, as also follows from definition (\ref{def}).  The chemical
potential is negative in the normal state and vanishes on approaching
the critical temperature $T_{\rm c}$, which is determined by the
condition 
\begin{equation} 
\label{condition}
n = G(0)\bigl|_{\mu =0}, 
\end{equation} 
where $n$ is the particle number density.  This equation is valid for
any $\mu<0$ and thus for any temperature $T>T_{\rm c}$.  However, it
starts to break down at the critical temperature where the condensate
starts to form in the $k=0$ state, which has bee ignored in Eq.\
(\ref{kint}).  The critical temperature is in other words the lowest
temperature where this equation is still valid.  From the criticality
condition it follows that upon approaching the critical temperature from
above, the chemical potential tends to zero in a $d$-dependent way as
$\mu(T) \sim -(T-T_{\rm c})^{2/(d-2)}$.  Below this temperature, in the
condensed state, the chemical potential remains zero.  The
$d$-dependence here distinguishes an ideal Bose gas from a Gaussian
theory, where $\mu$ tends to zero as $\mu(T) \sim |T-T_{\rm c}|$
irrespective of the dimensionality.  The critical exponents of an ideal
Bose gas, which can be extracted from the information given above, are
$\beta=1/2, \nu=1/(d-2)$, and $\gamma =2/(d-2)$.  As noted a long time
ago by Gunton and Buckingham \cite{GuntonBuckingham}, these exponents
are the same as for the spherical model in $d$ dimensions.  This model
corresponds to the limit $n \to \infty$ of the O($n$) spin model.

They continued to show that an ideal Bose gas with the modified energy
spectrum $\epsilon(k) \propto k^D$ is in the same universality class as
the spherical model with long-range interactions considered by Joyce
\cite{Joyce}.  The critical exponents for $D<d\leq2 D$, with $d=2D$ the
upper critical dimension, are given by
\begin{align} 
\label{ces}
\alpha & = \frac{d - 2 D}{d - D}, & \beta & = \frac{1}{2}, &
\gamma & = \frac{D}{d - D}, \nonumber \\ \nu & = \frac{1}{d - D}, &  \eta
& = 2 - D.
\end{align} 
The critical exponents of the standard ideal Bose gas are recovered by
setting $D=2$.  As will be shown shortly, the energy spectrum index $D$
equals the fractal dimension.

The partition function of the ideal Bose gas with a modified energy
spectrum can again be represented in the form (\ref{explicit}), with
$\tau = d/D +1$ and $\mu \propto (T - T_{\rm c})^{1/\sigma}$, where
$\sigma = 1/\nu D$ as for percolation [see Eqs.\ (\ref{nu}) and
(\ref{tau})]. The thermal wavelength $\lambda$ for the generalized Bose
gas is defined in a way that the spectrum can be written as
\begin{equation} 
\epsilon/k_{\rm B} T = c_D \lambda^D k^D,
\end{equation} 
with the constant $c_D$ conveniently chosen such that
\begin{equation} 
c_D^{d/D} = \frac{1}{(4 \pi)^{d/2}} \frac{2}{D}
\frac{\Gamma(d/D)}{\Gamma(d/2)},
\end{equation} 
where $\Gamma(x)$ is the Euler gamma function.  The correlation function
$G(x)$ at $x=0$ is again given by Eq.\ (\ref{G0}) while also the
integral of this function over space yields the same result as for $D=2$
given in Eq.\ (\ref{sumG}).  From the criticality condition
(\ref{condition}) it follows that the correlation length scales with the
chemical potential as $\xi \propto \mu^{-1/D}$, where---as already
indicated by our choice of notation---$D$ denotes the fractal dimension
as calculated from its definition (\ref{Hausdorff}).  The critical
exponents (\ref{ces}) now follow easily.

\begin{figure}
\begin{center}
\psfrag{d}[t][t][1][0]{$d$}
\psfrag{x}[t][t][1][0]{$x$}
\psfrag{t}[t][t][1][0]{$\tau$}
\psfrag{b1}[t][t][.8][0]{$\hbar /k_{\rm B} T$}
\includegraphics[width=6.cm]{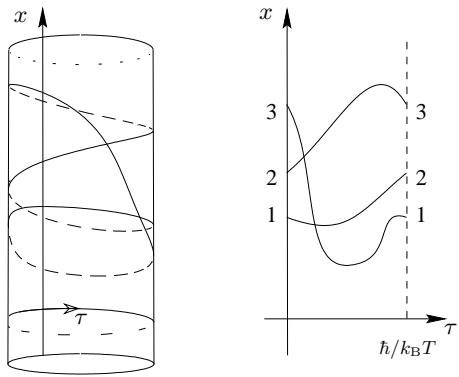}
\end{center}
\caption{A worldline loop involving three particles projected onto the
finite imaginary time interval $0 \leq \tau \leq \hbar/k_{\rm B} T$
(left panel).  The loop, which has winding number $w=3$, can be
equivalently represented as the worldlines of three cyclically permuted
particles (right panel). (After Ref. \cite{loops}.)
\protect\label{fig:path_labels}}
\end{figure}

\begin{figure}
\begin{center}
\psfrag{y}[t][t][1][0]{$y$} 
\psfrag{x}[t][t][1][0]{$x$}
\includegraphics[width=6.cm]{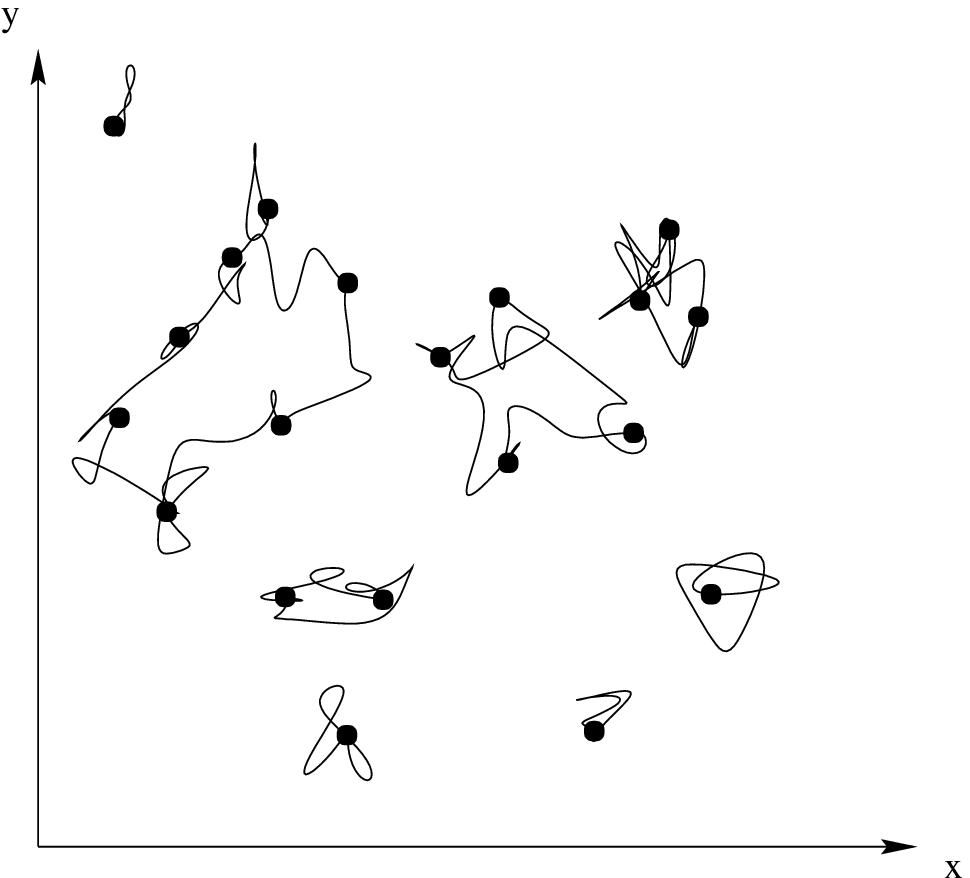} \\
\includegraphics[width=6.cm]{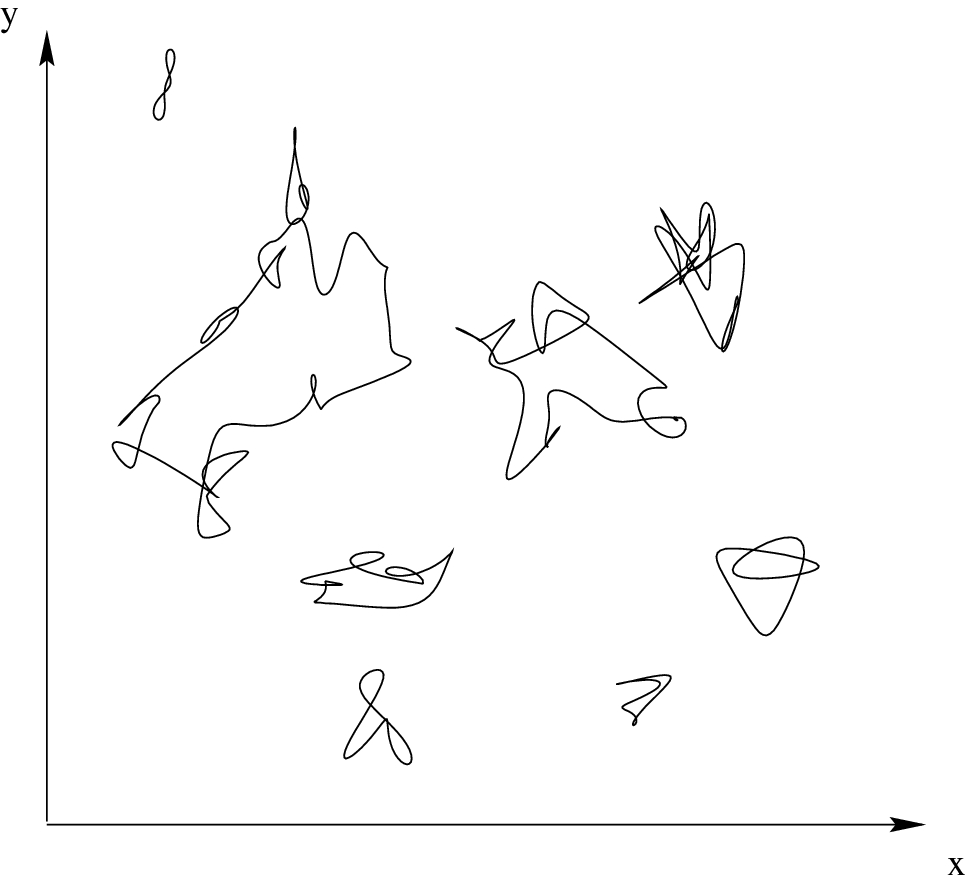}
\end{center}
\caption{Worldline loops of $20$ particles in two space dimensions.
Some of the worldlines are grouped together in single rings.  The beats
in the top panel represent particles moving in imaginary time along the
strings towards their own initial position, in case of a single particle
in a ring, or towards that of the neighboring particle in front of them.
After an imaginary time $\hbar/k_{\rm B} T$, these positions are reached
and the particles cyclically permuted.  In the bottom panel the beads
are omitted, leaving behind a loop gas of worldlines.  (After
Ref. \cite{loops}.)  \protect\label{fig:dots}}
\end{figure}

As detailed in Ref.\ \cite{loops}, the fugacity series (\ref{explicit})
represents a sum over loops with the distribution (\ref{loopdis}), or,
equivalently,
\begin{equation}
\label{loopdisgen} 
l_w(T)  = (\ell/a)^{- \tau} \exp(-\theta \ell/k_{\rm B} T), \quad
\theta \propto (T-T_{\rm c})^{1/\sigma},
\end{equation} 
with $\ell = w a$ the ``length'' measured in units of the characteristic
``length'' scale $a =\hbar \beta$, and $\theta = - \mu k_{\rm B}
T/\hbar$ the string ``tension''.  The first factor in (\ref{loopdisgen})
measures the configurational entropy of loops, while the second is a
Boltzmann factor.  Each loop is characterized by the winding number $w$,
telling how often it wraps around the imaginary time axis.  Physically,
a loop represents the worldlines in imaginary time of $w$ particles
grouped together in a single ring known from Feynman's theory of the
$\lambda$-transition in $^4$He \cite{Feynman53}.  Particles in the same
ring are cyclically permuted after an imaginary time $\hbar/k_{\rm B} T$
(see Fig.\ \ref{fig:path_labels} for an illustration involving three
particles).  With increasing imaginary time, the particles move along
closed strings towards the initial position of the particle in front of
them (see Fig.\ \ref{fig:dots}).  When only one particle is contained in
a ring, the particle returns to its own initial position after an
imaginary time $\hbar/k_{\rm B} T$.  On approaching the critical
temperature from above, the chemical potential and thus the string
tension become smaller, and loops with larger winding numbers start to
appear.  At $T_{\rm c}$, the string tension vanishes and long loops
containing arbitrarily many cyclically permuted particles appear in the
system, signalling the onset of Bose-Einstein condensation.  Above the
critical temperature, long loops are exponentially suppressed.  Because
the loops are worldlines embedded in spacetime and parameterized by
Euclidean time, $\ell$ and $a$ have the dimension of time, not of
length---whence the quotation marks used below Eq.\ (\ref{loopdisgen}).

From this and also from comparing the partition function
(\ref{explicit}) of an ideal Bose gas with the one for percolation, Eq.\
(\ref{perZ}), we observe a close connection between the two phenomena.
The rings of cyclically permuted particles in a Bose gas correspond to
the clusters of percolation theory.  The number of Bose particles
contained in a ring corresponds to the number of sites contained in a
cluster.  In the same way that clusters grow when the percolation
threshold is approached from below, rings with larger winding numbers
appear in a Bose gas on approaching the critical point from above.  In
both cases, the behavior at the critical point is algebraic, rather than
exponential.  The percolating cluster at $p>p_{\rm c}$ corresponds to
the Bose-Einstein condensate at $T<T_{\rm c}$.  [The reason for ignoring
the $k=0$ contribution in Eq.\ (\ref{kint}) was to bring out this
analogy as clearly as possible.]  But there are also differences.

Although the exponents (\ref{ces}) satisfy the scaling relations,
$\beta$, $\gamma$, and $\eta$ are not related to the exponents $\sigma$
and $\tau$ parameterizing the loop distribution (\ref{loopdis}) in the
same way as in percolation theory [see Eq.\ (\ref{perce})].  Using these
equations with the ideal Bose gas values for $\tau$ and $\sigma$, we
find instead
\begin{equation} 
\beta_{\rm per} = 1, \quad \gamma_{\rm per} = - \frac{d-2 D}{d-D}, \quad
\eta_{\rm per} = 2 + d - 2 D,
\end{equation} 
where we have given the exponents also the subscript ``per'' to indicate
that we used the percolation formulas for them.  

Despite the differences in form, these exponents too satisfy the scaling
relations $2 \beta_{\rm per} + \gamma_{\rm per} = 2 - \alpha$ and
$\gamma_{\rm per} = \nu (2 - \eta_{\rm per})$, with $\alpha$ and $\nu$
given in (\ref{ces}).  Apparently, there are two different sets of
critical exponents, both satisfying the scaling laws.

To understand this, let us compare the sum rules (\ref{sumGper}) and
(\ref{sumG}) satisfied by the correlation function of percolation and an
ideal Bose gas, respectively.  There is no obvious connection between
the two, implying that the ``susceptibility'' and also the
``magnetization'' are defined differently in the two systems.  In
percolation theory, they were defined in terms of the partition function
in Eq.\ (\ref{PS}).  With $c$ replaced by $-\mu/k_{\rm B} T$ to map the
cluster distribution of percolation theory onto the loop distribution of
an ideal Bose gas, it follows that $P$ and $S$ correspond to the
condensate density $n_0$ and the compressibility $\kappa$ of
an ideal Bose gas, respectively, with
\begin{equation} 
n_0 \sim (T_{\rm c} - T)^{\beta_{\rm per}}, \quad  \kappa \sim |T_{\rm
c} - T|^{-\gamma_{\rm per}}.
\end{equation}
The exponents $\beta$ and $\gamma$ featuring in Eq.\ (\ref{ces}) are, on
the other hand, connected to the order parameter of an ideal Bose gas,
which is the field $\phi_0$ describing the condensate, with $|\phi_0|^2
= n_0$.  This last equation relates the two $\beta$-exponents, giving
$\beta_{\rm per} = 2 \beta$.  Since $\beta_{\rm per} = 1$, it also
follows that $\gamma_{\rm per} = - \alpha$, as one would expect naively.
It should be noted that the exponents $\nu$ and $\alpha$ are not
sensitive to the nature of the order parameter describing the
transition.  Given the presence of a single diverging length scale, any
sensible definition of the correlation length leads to the same exponent
$\nu$.  And $\alpha$ merely characterizes the algebraic behavior of the
logarithm of the partition function, which is essentially the pressure.
Despite the difference in the correlation function of percolation and an
ideal Bose gas, the relation (\ref{nu}) between the correlation length
exponent $\nu$ and the fractal dimension is the same for the two, as can
be easily checked repeating the derivation of this relation with the
correlation function of an ideal Bose gas.

A more fundamental difference between percolation theory and the string
picture of an ideal Bose gas is that in the latter, the number of
particles is fixed, while in the former there is no condition on the
number of occupied sites.  The fixed number of particles in an ideal
Bose led to the criticality condition (\ref{condition}) with $G(0)$
given in Eq.\ (\ref{G0}) which in turn determined the critical exponent
$\nu$.  In percolation theory, where there is no such a condition, $\nu$
is not determined that way.  As a result, although the two phenomena are
described by essentially the same theory specified by a loop or cluster
distribution, they in general are not in the same universality class.
Only the special case of percolation in the upper critical dimension
$d_{\rm u}=6$, where $D=4$, or on a Bethe lattice is in the same
universality class as an ideal 6-dimensional Bose gas with fractal
dimension $D=4$.  

However, in the absence of a criticality condition, we expect
uncorrelated percolation and strings of the type appearing in the
description of an ideal Bose gas to be described by exactly the same
theory.  This is indeed what has been observed in numerical simulations
on random string networks \cite{StroblHindmarsh}.

\section{Random String Networks}
\label{sec:String}
These simulations describe theories with a global U(1) gauge symmetry,
which undergo a phase transition from a symmetric state to one where the
U(1) symmetry is spontaneously broken.  The ordered state, which is
characterized by a complex order parameter, can have topological line
defects---known as cosmic strings in the context of cosmology, and as
vortices in the context of condensed matter.  The line defects are
either closed or terminate at the boundary, they cannot terminate inside
the system.  Upon circling a vortex of unit strength, the phase
$\vartheta$ of the order parameter changes continuously by $2 \pi$.

To numerically study line defects, Vachaspati and Vilenkin
\cite{VachaspatiVilenkin} considered a cubic lattice with periodic
boundary conditions.  They discretized the vacuum manifold of the
ordered state by allowing $\vartheta$ to take on only certain discrete
values in the interval $0 \leq \vartheta < 2 \pi$.  The restriction to
this interval assures that only vortices of unit strength are generated.
Each lattice site is assigned one of the discrete values at random.
Phases at different lattice sites are therefore uncorrelated.  Vortices
are traced by going around each plaquette of the lattice once.  If a
phase difference of $2\pi$ is found upon returning to the site one
started at, it is concluded that a vortex penetrates the plaquette.  In
going from one lattice site to a neighboring one, the geodesic rule
introduced by Kibble \cite{Kibble} is implemented, assuming that the
(discrete) changes in $\vartheta$ are minimal.  When two instead of just
one vortex is found to penetrate a unit cell, the two incoming and two
outcoming vortex segments are randomly connected.  There are different
ways to do this, the simplest one being to choose equal probabilities as
was done in Ref.\ \cite{VachaspatiVilenkin}.  But there are other
choices, such as favoring those possibilities that generate longer loops
\cite{Kajantieetal}.  In this way a whole random network of vortices (of
unit strength) can be traced out.

Such a random string network is of relevance to the formation of
topological defects after a rapid phase transition involving the
spontaneous breakdown of some symmetry \cite{leshouches}.  As was first
proposed by Kibble \cite{Kibble} in the context of cosmology, causally
disconnected pieces of the early universe may after a rapid temperature
quench end up in different, but degenerate states of the broken-symmetry
vacuum, characterized by different values for the phase $\varphi$ of the
order parameter.  When these regions come together, the resulting
frustration at the boundaries may lead to the formation of topological
defects.  Zurek \cite{Zurek} noted that a similar mechanism might
operate in the context of condensed-matter physics and developed a
qualitative theory, based on scaling arguments, of the formation of
topological defects in time-dependent phase transitions.  To test this
scenario, superfluid $^3$He-B in rotation was radiated with neutrons
\cite{Helsinki}.  After a neutron-absorption event, in which a small
region is briefly heated up to the normal phase, a vortex network was
produced.  Under the action of the Magnus force, some of the vortex
loops in the network expand and connect to the container wall.  They are
then, still under the action of the Magnus force, pulled straight and,
finally, end up in the center of the cylindrical container aligned
parallel to the rotation axis, where they can be detected by means of
NMR.  The number of vortices produced in the neutron irradiation
experiments was in agreement with Zurek's predictions.  The vortex
network was numerically simulated \cite{Eltsovetal} with the
Vachaspati-Vilenkin algorithm discussed above.

Besides finite vortex loops, also infinite vortices appear in a random
network with high enough vortex densities.  The division between finite
vortex loops and infinite ones is of course somewhat arbitrary on a
finite lattice.  A possible choice is to classify vortices with a length
larger than $2L^2$, where $L$ is the lattice size, as infinite.
Universal quantities such as the critical exponents should be
independent of the precise choice.  The fractal dimension found in the
initial simulations \cite{VachaspatiVilenkin} were consistent with that
of a Brownian random walk, $D=2$.  This result can be understood
\cite{HindmarshStrobl} by realizing that a dense random network was
simulated, where a single vortex experiences an effective repulsion from
the neighboring vortex segments as the volume occupied by these segments
is not available anymore.  The situation is similar to a polymer in a
dense solution, which also displays the structure of a Brownian random
walk.

Without any external field, all phase values of the order parameter have
equal probability.  Vachaspati \cite{Vachaspati} lifted this degeneracy
of the vacuum manifold by giving some values a bias, thus obstructing
the formation of vortices and decreasing the vortex density.  When
studying the statistics of the random network as function of the bias, a
threshold was found, above which infinite vortices are absent and finite
loops exponentially suppressed.  The sudden appearance of infinite
vortices as the threshold is approached from the low-density side, was
interpreted as a percolation process.

Strobl and Hindmarsh \cite{StroblHindmarsh} carried out more extensive
numerical simulations on the transition of a random string network
triggered by changing the bias.  Using the percolation set of critical
exponents, they measured---to within statistical errors---values similar
to those for 3-dimensional uncorrelated site percolation, which has
$\sigma=0.45$, $\tau=2.18$ and $D=2.53$ \cite{StauferAharony}.  For the
vortex loops, the value $D>2$ means that they are self-seeking.  The
conclusion that the phase transition of a random string network is
identical to the uncorrelated percolation transition is precisely what
one would expect given our observations of the last section.

\section{Correlated Percolation}
\label{corper}
Up to this point we considered only {\it uncorrelated} site percolation,
where lattice sites were occupied at random, and nearest-neighboring
occupied sites grouped together in clusters.  However, as is known from
work on the Ising model \cite{StauferAharony}, this is not the whole
story.

In the context of the Ising model, it is more appropriate to refer to
occupied sites as sites with spin up, say.  In general, the percolation
threshold temperature is below the critical one.  That is to say,
already before the thermal phase transition is reached, clusters of
spin-up sites percolate the lattice.  Only on a 2-dimensional square
lattice, the site percolation threshold temperature coincides with the
critical temperature.  Despite this, the Ising-model and percolation
critical exponents are different--as in higher dimensions.  The
situation is similar to the one discussed at the end of Sec.\
\ref{sec:BEC}, with the percolation transition being in a different
universality class as the thermal one.

It is nevertheless possible to modify the definition of clusters in such
a way that the critical behavior of the modified clusters becomes
identical to that of the Ising model \cite{CK}.  This approach goes back
to seminal work by Fortuin and Kasteleyn \cite{FortuinKasteleyn}, who
mapped the partition function of the Ising model onto a {\it correlated}
percolation problem.  To obtain a modified cluster, take two
nearest-neighboring sites $i, j$ of a standard cluster with all the
spins up, and add the pair to the new cluster with the bond probability
$p_{i j} = 1 - \exp(-2 J/k_{\rm B} T)$, where $J$ is the spin-spin
coupling of the Ising model.  The factor $2J$ appearing here corresponds
to the increase in energy when one of the two spins involved in the bond
is flipped.  The upshot is that the bond clusters are in general smaller
than the standard ones and also more loosely connected.  The bond
clusters percolate right at the critical temperature and display
critical behavior identical to that of the Ising model.

The method has been turned into a powerful Monte Carlo algorithm by
Swendson and Wang\cite{SwendsonWang}, and by Wolff \cite{Wolff}, where
not individual spins are updated, but entire clusters.  The main
advantage of the nonlocal cluster algorithm in comparison to a local
update algorithm, like Metropolis or heat bath, is that it drastically
reduces the critical slowing down near the critical point.

Very recently, also the critical behavior of the 3-dimensional O($n$)
spin models, with $n=2,3,4$, was described as a correlated percolation
process \cite{Satzetal}.  The relevant clusters percolate at the
critical temperature, and were shown to have the same critical exponents
as the corresponding models.

\section{Worldline Loop Gases}
\label{sec:LG}
In the same way that correlated percolation can describe thermal phase
transitions of interacting systems also correlated worldline loop gases
can.  Recall that the loop gas description of an ideal Bose gas hinged
on the representation (\ref{explicit}) of the partition function in
terms of the loop distribution (\ref{loopdisgen}), which derived from
the representation (\ref{kint}) as an integral over momentum.  If we
assume that near the critical point, the partition function of an
interacting system can be put in a similar form, but with $E(k)$
replaced by the single-particle spectrum of the excitation that becomes
critical at $T_{\rm c}$, the integral can in principle be done, yielding
some loop distribution $l_w$, with $w$ physically denoting the number of
times a worldline loop wraps around the imaginary time axis.  As for the
ideal Bose gas, this distribution is parametrized by two exponents
$\sigma$ and $\tau$, and can be put in the general form
(\ref{loopdisgen}), with values characteristic for the universality
class under consideration.  The string tension of the worldline loops
vanishes at $T_{\rm c}$, and loops with arbitrarily large winding number
$w$ appear in the system, signalling the onset of Bose-Einstein
condensation in the low-temperature phase.  Note that the representation
(\ref{kint}) of the partition function as an integral over momentum does
not require a nonrelativistic theory.  Also relativistic theories lead
to a similar form for the (equilibrium) partition function at finite
temperature \cite{Kapusta}, as it is essentially fixed by Bose
statistics.

The generalization of the sum rule (\ref{sumG}) to correlated worldline
loop gases reads
\begin{equation} 
\label{sumGgen}
\int \dd^d x \, G(x) = \sum_w {\rm e}^{- \theta \ell /k_{\rm B} T},
\end{equation} 
where we recall that the string tension behaves near the critical point
as $\theta \propto (T-T_{\rm c})^{1/\sigma}$.  The sum rule leads
directly to the relation
\begin{equation} 
\label{gamma}
\gamma = 1/\sigma,
\end{equation} 
while at criticality, with the long distance cutoff $x_{\rm max}= L$,
corresponding to $w_{\rm max} = L^D$ as before, it gives the relation
\begin{equation}
\label{eta} 
\eta = 2 - D.
\end{equation} 
These two relations generalize the corresponding results in Eq.\
(\ref{ces}) for an ideal Bose gas with modified energy spectrum.  The
value $\eta=0$ separates self-avoiding worldlines ($D<2$) from
self-seeking ones.  Since the fractal dimension has to be positive and
smaller than the dimension $d$ of the embedding space \cite{Hoveetal},
$2-d < \eta < 2$.  Translated in terms of limits on $\tau$, it follows
that $2 < \tau < d/2 + 1$ for self-seeking worldlines, while $\tau > d/2
+ 1$ for self-avoiding ones.  Note that in two space dimensions and
below, no self-seeking strings are possible.

As in Sec.\ \ref{sec:BEC}, the exponents here are those for describing
the thermal phase transition of a loop gas of strings (worldlines).
They are to be distinguished from the ones defined in the context of
percolation theory given in (\ref{perce}).  The two sets have, however,
the exponents $\alpha$ and $\nu$ in common, implying that the present
set can also be expressed entirely in terms of the exponents $\sigma$
and $\tau$ parameterizing the loop distribution (\ref{loopdisgen}).
Indeed, we find
\begin{align} 
\label{stringce}
\alpha & = 2 - \frac{\tau -1}{\sigma}, & \beta &= \frac{\tau
-2}{2\sigma}, & \gamma & = \frac{1}{\sigma}, \nonumber \\ \eta
&= 2 - \frac{d}{\tau-1}, & \nu & = \frac{\tau -1}{d \sigma}, & D & =
\frac{d}{\tau-1}.
\end{align} 
Since only general scaling laws are used in deriving these results, we
believe they hold for any critical theory specified by a loop
distribution (\ref{loopdisgen}) with exponents $\sigma$ and $\tau$.  It
can be easily verified that for an ideal Bose gas with $\sigma = d/D -1$
and $\tau = d/D +1$, Eq.\ (\ref{stringce}) reduces to the correct values
(\ref{ces}).

As an illustration, let us discuss the worldline loop distribution of
some well-known statistical models.  We start with random walks.
\begin{table}
\begin{tabular}{l|ll|llllll}
\hline \hline & & & & & & & & \\[-.2cm]
Random walk & $\sigma$ & $\tau$ & $\alpha$ & $\beta$ & $\gamma$ &
$\eta$ & $\nu$ & $D$ \\[.2cm]  \hline  & & & & & & & & \\[-.2cm] 
Brownian & $1$ & $1+d/2$ & $2-d/2$ & $d/4-1/2$ & $1$ & $0$ & $1/2$ & $2$
\\[.2cm]
Smooth & $1$ & $1+d$ & $2-d$ & $d/2-1/2$ & $1$ & $1$ & $1$ &
$1$ \\[.2cm] \hline \hline 
\end{tabular}
\caption{Critical exponents for Brownian and smooth random walks,
together with the exponents $\sigma$ and $\tau$ parameterizing the loop
distribution (\ref{loopdisgen}).  \protect\label{table:RW}}
\end{table}
In Table \ref{table:RW}, the critical exponents of two different types
of random walks are given, together with the corresponding values for
$\sigma$ and $\tau$ parameterizing the loop distribution
(\ref{loopdisgen}).  Both have the value $\sigma =1$ in common.  The
first one is the Brownian random walk \cite{KleinertPath}, which has a
fractal dimension $D=2$, while the second one is the so-called smooth
random walk \cite{Polyakov,Ambjorn}, which has $D=1$ and is therefore
self-avoiding.  Whereas a Brownian random walk can be understood as a
particle hopping through a lattice with its next (nearest-neighboring)
site chosen randomly at each step, a smooth random walk can be
understood as a particle hopping through the lattice with its velocity
rather than its position being changed randomly at each step.  As a
result, whereas a typical path of a Brownian random walk is continuous
but not differentiable, a typical path of a smooth random walk is
differentiable while only the first derivative is not.  The reluctance
to change directions is typical for fermionic particles.  To understand
the value $\eta=1$ for a smooth random walk, note that the correlation
function of a Brownian random walk is given in the continuum by,
\begin{equation} 
G(x) = \int \frac{\dd^d k}{(2 \pi)^d} \, G_{\rm B}(k) \, {\rm e}^{i {\bf
k} \cdot {\bf x}} ,
\end{equation}
with
\begin{equation} 
G_{\rm B} (k) = \frac{1}{k^2 + \xi^{-2}},
\end{equation} 
leading to the algebraic behavior (\ref{algebraic}) at criticality with
$\eta = 0$.  A smooth random walk, on the other hand, has a correlation
function involving one factor of momentum less \cite{Polyakov,Ambjorn},
\begin{equation} 
G_{\rm S}(k) = \frac{1}{{\bf k} \cdot {\bf e} + \xi^{-1}},
\end{equation} 
where ${\bf e}$ is some unit vector.  At criticality, this again gives
the algebraic behavior (\ref{algebraic}), but now with $\eta=1$.

In two space dimensions, the critical exponents of various models are
known exactly.  For example, for O($n$) spin models with $n = -2 \cos(2
\pi/t)$ and $1 \leq t \leq 2$ they are given by \cite{Nienhuis},
\begin{align} 
\label{os}
\alpha & =\frac{2 t -3 }{t - 2} , & \beta & = \frac{(t-1) (t-3) }{8t(t
-2)} , & \gamma & = - \frac{t^2 + 3}{4t(t-2)}, \nonumber \\ \nu & =
- \frac{1}{2(t-2)}, & \eta & = - \frac{(t-1) (t-3) }{2t} , & D & =
\frac{t^2 + 3}{2t},
\end{align}  
corresponding to 
\begin{equation} 
\sigma = - \frac{4t(t - 2)}{t^2 + 3}, \quad \tau = \frac{(t + 1) (t +
3)}{t^2 + 3}.
\end{equation} 
For special values of the parameter $t$ such that $n$ is an integer, one
obtains the critical exponents for the Gaussian model, or the Brownian
random walk ($n=-2$), the self-avoiding walk (SAW) ($n=0$), the Ising
model ($n=1$), and the XY-model ($n=2$).  The critical exponents for
these models together with those of the spherical model are collected in
Table \ref{table:On}.
\begin{table}
\begin{tabular}{l|l|ll|llllll}
\hline \hline & & & & & & & & & \\[-.2cm] 
Model & $n$ & $\sigma$ & $\tau$ & $\alpha$ & $\beta$ & $\gamma$ & $\eta$
& $\nu$ & $D$ \\[.2cm]
\hline & & & & & & & & & \\[-.2cm]
Gaussian & $-2$ & $1$ & $2$ & $1$ & $0$ & $1$ & $0$ & $1/2$ & $2$
\\[.2cm] 
SAW & $0$ & $32/43$ & $91/43$ & $1/2$ & $5/64$ & $43/32$ & $5/24$ &
$3/4$ & $43/24$ \\[.2cm]
Ising & $1$ & $4/7$ & $15/7$ & $0$ & $1/8$ & $7/4$ & $1/4$ & $1$ & $7/4$
\\[.2cm]
XY & $2$ & $0$ & $15/7$ & & & & $1/4$ & & $7/4$ \\[.2cm] \hline & & & &
& & & & \\[-.2cm]
Spherical & $\infty$ & $0$ & $2$ & & $1/2$ & & $0$ & & $2$ \\[.2cm]
\hline \hline
\end{tabular}
\caption{Critical exponents for the 2-dimensional O($n$) spin model,
with $n=-2,0,1,2, \infty$, respectively, together with the exponents
$\sigma$ and $\tau$ parameterizing the loop distribution
(\ref{loopdisgen}).  \protect\label{table:On}}
\end{table}

A few points may be worth noting.  (i) Both the Ising ($t=3/2$)
and the XY-model ($t=2$) have the same fractal dimension.  This is made
possible by a minimum in $D(t)$ at $t=\sqrt{3}$ located between the two
models, where $D=\sqrt{3}$.

(ii) In the literature, the fractal dimension is often equated to
$1/\nu$ rather than to $1/\sigma \nu$ as in Eq.\ (\ref{nu}).  We
believe, the result (\ref{nu}) is more general.

(iii) The special character of the 2-dimensional XY and also of the
spherical model is seen in the loop gas picture by the vanishing of the
exponent $\sigma$, implying the absence of the Boltzmann factor in the
loop distribution function (\ref{loopdisgen}).

(iv) Both the 2-dimensional Ising and XY-model share the same
$\tau$-value and their worldlines have, consequently, the same fractal
dimension.  There is at least one other 1-dimensional object embedded in
two space dimensions with the same fractal dimension.  Surprisingly, it
is the perimeter of a percolation cluster of standard, uncorrelated
percolation in two dimensions \cite{Sapovaletal}, which can also be
mapped onto various types of random walks.  Close to criticality, the
perimeter of the percolating cluster is again distributed according to
Eq.\ (\ref{perdis}), but with different values for the exponents.  To
avoid confusing with the 2-dimensional cluster exponents $\sigma =
36/91$ and $\tau = 187/91$ corresponding to $D=91/48$, we call the
perimeter exponents $\sigma'$ and $\tau'$.  Since there is only one
correlation length in the system, the cluster and perimeter exponents
satisfy \cite{WeinribTrugman}
\begin{equation} 
\label{link}
(\tau-1)/\sigma = (\tau'-1)/\sigma', \quad \sigma/\sigma' = D'/D.
\end{equation} 
The perimeter values \cite{Ziff}
\begin{equation} 
\sigma' = 3/7, \quad \tau' = 15/7,
\end{equation} 
leading to $D'=7/4$ indeed fulfill the relations (\ref{link}).

Recent high-precision Monte Carlo studies of the 3-dimensional
XY-universality class gave \cite{HasenbuschToeroek}
\begin{equation}
\label{numm}
\nu=0.6723(3), \quad \eta=0.0381(2).
\end{equation} 
The resulting exponents parameterizing the worldline loop distribution
are
\begin{equation} 
\sigma = 0.7582,  \quad \tau = 2.5291,
\end{equation} 
corresponding to a fractal dimension $D = 1.9619$ very close to two, in
accordance with the smallness of $\eta$.  Since $D<2$, the worldline
loops of the 3-dimensional XY-model are self-avoiding.

\section{Vortex Loop Gases}
\label{sec:VL}

Let us next turn to interacting systems with a global U(1) gauge
symmetry, which have vortices as topological defects in the ordered
state.  Thermal phase transitions in systems of this type have, besides
the conventional Landau description, an alternative description in terms
of proliferating vortices \cite{GFCM}.  Such a scenario was first put
forward by Onsager \cite{Onsager} and later by Feynman \cite{Feynman55}
to describe the $\lambda$-transition in $^4$He.  In the ordered,
superfluid state, where the U(1) symmetry is spontaneously broken, only
finite vortex loops are present.  Because of the finite string tension,
the loops are exponentially suppressed.  Upon approaching the critical
temperature from below, the string tension vanishes, and vortices start
to proliferate.  For vanishing string tension, vortices can gain
configurational entropy by growing without energy cost.  The resulting
infinite vortices disorder the system and thereby restore the
spontaneously broken U(1) symmetry.  The importance of vortex loop
excitations in triggering the phase transition has since then been
emphasized by various authors \cite{70ies}.  Early numerical evidence
for this picture based on Monte Carlo simulations of the 3-dimensional
XY-model was given by Janke and Kleinert \cite{Janke,GFCM}.  Analytic
methods to describe the transition using vortex loops in a qualitative
way were developed by Williams \cite{Williams}, and by Shenoy and
collaborators \cite{Shenoyetal}.

Like worldline loops, vortex loops are specified by a loop distribution
as in Eq.\ (\ref{loopdisgen}) parameterized by two exponents $\sigma'$
and $\tau'$ \cite{Coplandetal}.  These determine the critical
exponents of the system again through Eq.\ (\ref{stringce}), with
$\sigma$ and $\tau$ replaced by $\sigma'$ and $\tau'$.  The result that
the exponent $1/\sigma'$ parameterizing the vanishing of the vortex
string tension directly gives the critical exponent $\gamma$---and not
$2 \nu$ as is often claimed in the literature, also by the present
author \cite{loops,leshoucheslect}---was first observed by Nguyen and
Sudb\o \cite{NguyenSudbo}, while the connection $\eta = 2 - D$ for
vortex loops was very recently derived independently in Ref.\
\cite{Hoveetal}.

In the past few years various groups have continued to numerically
investigated the 3-dimensional XY-model from the perspective of
vortices, in particular their loop distribution
\cite{AntunesBettencourt,NguyenSudbo,Kajantieetal}.  Let us discuss the
numerical values for the exponents $\sigma'$ and $\tau'$ obtained in
these studies from the present perspective, concentrating on insights
not available before.

Irrespective of the description chosen, either with the help of
worldline loops or vortex loops, the expression for the correlation
length exponent is formally the same.  Since also the numerical value of
this critical exponent is independent of the description, the exponents
$\sigma$ and $\tau$ parameterizing the worldline loop distribution and
$\sigma'$ and $\tau'$ parameterizing the vortex loop distribution
satisfy $(\tau-1)/\sigma = (\tau'-1)/\sigma'$ as in Eq.\ (\ref{link}).
They should both lead to the same numerical value for $\nu$ given in
Eq.\ (\ref{numm}).  Using for $\sigma'$ and $\tau'$ the numerical values
\cite{note}
\begin{equation} 
1/\sigma' = 1.45(5), \quad \tau' = 2.4(1)
\end{equation} 
reported by Nguyen and Sudb\o \cite{NguyenSudbo}, we find from Eq.\
(\ref{stringce}) a result $\nu = 0.68$ very close to the expected one
(\ref{numm}).  Since $2 < \tau' < d/2 + 1$, vortex loops in the XY-model
are (slightly) self-seeking.  The numerical values for $\sigma'$ and
$\tau'$ reported by the other groups give the results $\nu = 0.62$
\cite{AntunesBettencourt} and $\nu = 0.60$ \cite{Kajantieetal}, which
are within about $10\%$ of the expected result.

It was noted by Williams \cite{Williams} that the fractal dimension $D$
extracted from the value $\tau'=2.23(4)$ reported in Ref.\
\cite{AntunesBettencourt} was close to the estimate $D=(d+2)/2$ obtained
from a Flory-type of argument \cite{Shenoyetal}.  This estimate is
identical to de Gennes' one (\ref{deGennes}) for uncorrelated
percolation which was also based on a Flory-type argument.  It should be
noted, however, that de Gennes' modification of the original Flory
argument was motivated in part to arrive at the correct upper critical
dimension $d_{\rm u}=6$ for percolation, rather than $d_{\rm u}=4$ for
random walks and the XY-model.

Kajantie and collaborators \cite{Kajantieetal} investigated different
algorithms to construct the string network.  Depending on the algorithm
chosen, they obtained (slightly) differing quantitative results.  None
of the algorithms used yielded a percolation threshold precisely at the
critical temperature.  From this observation they concluded that
``geometrically defined percolation observables need not display
universal properties related to the critical behavior of the system''.
It has been suggested \cite{private} that this discrepancy might instead
hint at the existence of a unique, yet to this date unknown, algorithm to
construct the string network that would give precisely the expected XY
critical behavior.  The situation would then be similar to that found
for correlated percolation and thermal phase transitions in O($n$)
spin models discussed in Sec.\ \ref{corper}.

\section{Conclusions}
\label{sec:concl}
In this paper we discussed the close analogy between percolation of
(correlated) clusters and proliferation of strings at thermal phase
transitions of interacting systems.  Two types of strings were
considered: worldlines and topological line defects.  Both are described
in a way similar to clusters in percolation theory, being specified
close to the critical point by a distribution parametrized by only two
exponents.  At the critical temperature, the string tension vanishes and
loops proliferate in the same way as clusters percolate.  If the loops
are finite-temperature worldlines, their proliferation signals the
formation of a Bose-Einstein condensate which spontaneously breaks the
global symmetry characteristic of the phase transition under
consideration.  If, on the other hand, the loops are vortices, their
proliferation signals the disordering of the ordered state in which
these vortices are topologically stable.

In deriving our results we used the ideal Bose gas with modified energy
spectrum as a stepping stone.  The model, despite being noninteracting,
has nontrivial critical exponents which are known exactly.  It also has
the virtue that it can be mapped onto a loop gas in an exact way.  The
resulting relations between the exponents parameterizing the loop
distribution and the critical exponents describing the thermal phase
transition of the system were shown to be easily generalized to
interacting loop gases using general scaling laws.  As for percolation,
the two exponents parameterizing the distribution were shown to encode
the entire set of critical exponents.

Various statistical models with exactly known critical exponents were
discussed from the perspective of a worldline loop gas.  Also recent
numerical studies on the statistical properties of the vortex loop gas
in the 3-dimensional XY-model were discussed.

It was shown that a random string network is to thermal phase
transitions involving correlated strings as what uncorrelated site
percolation is to thermal phase transitions in O($n$) spin models.

A possible connection of the approach discussed in this paper with the
Monte Carlo loop algorithm introduced by Everetz \cite{Everetz} is
presently under investigation.

\acknowledgments I would like to thank M. Krusius for the kind
hospitality at the Low Temperature Laboratory in Helsinki, K. Kajantie,
T. Neuhaus, and A. Nguyen for helpful discussions, H. Satz for inspiring
lectures at the Zakopane summer school \cite{Satz}, T. Vachaspati and
G. Williams for useful correspondence.

This work was funded by the EU sponsored programme Transfer and Mobility
of Researchers under contract No.\ ERBFMGECT980122. 

The work is performed as part of a scientific network supported by the
ESF (see the network's URL,
http://www.physik.fu-berlin.de/$\sim$defect).
%
%

%

%
%\end{multicols}

\begin{thebibliography}{99}
\bibitem{Flory} The original Flory argument for self-avoiding random
walks gave $D_{\rm Fl} = (d+2)/3$.  For space dimensions lower than the
upper critical dimension, which is $d_{\rm u}=4$ for a self-avoiding
random walk, this estimate gives a fractal dimension smaller than that
for a Brownian random walk $D_{\rm Fl} < 2$.  A fractal dimension $D>2$
corresponds to a self-seeking random walk.
\bibitem{note} The value for $\tau'$ we use was not reported in the main
text of Ref.\ \cite{NguyenSudbo}, but in footnote 74 of that reference.
\end{thebibliography}
\end{document}